\documentstyle[epsf,cite,12pt]{article}
\newcommand{\be}{\begin{equation}}
\newcommand{\ba}{\begin{eqnarray}}
\newcommand{\ee}{\end{equation}}
\newcommand{\ea}{\end{eqnarray}}
\newcommand{\nn}{\nonumber}

\newcommand{\GeV}{\;\mbox{GeV}}

\newcommand{\simlt}{\stackrel{<}{{}_\sim}}
\newcommand{\simgt}{\stackrel{>}{{}_\sim}}
\newcommand{\ol}{\overline}

\renewcommand{\i}{{\mathrm i}}

\newcounter{currequation}
\newenvironment{subeqns}{\setcounter{currequation}{\value{equation}}%
\stepcounter{currequation}\setcounter{equation}{0}%
\renewcommand{\theequation}{\arabic{currequation}\alph{equation}}%
\begin{eqnarray}}{\end{eqnarray}{\setcounter{equation}{\value{currequation}}}%
\renewcommand{\theequation}{\arabic{equation}}}
\begin{document}
\title{The Inflaton As Dark Matter}
\date{December 13, 1997}
\author{Saul Barshay and Georg Kreyerhoff\\
III.~Physikalisches Institut A\\
RWTH Aachen\\
D-52056 Aachen}
\maketitle
\vspace{-9cm}
\flushright {PITHA 97/46}
\vspace{8cm}\par
\centerline {\small (to be published in Zeitschr. f\"ur Physik C)} 
\abstract{Within the framework of an explicit dynamical model, in
which we calculate the radiatively-corrected, tree-level potential
that sets up inflation, we show that the inflaton can be a significant
part of dark matter today. We exhibit potentials with both a maximum and
a minimum. Using the calculated position of the potential minimum,
and an estimate for fluctuations of the inflaton field in the early
universe, we calculate a contribution to the matter energy density
of $(1-2)\times 10^{-47}\GeV^4$ in the present universe, from cold inflatons
with mass of about $6\times 10^9\GeV$. We show that the inflaton might
decay in a specific way, and we calculate a possible lifetime 
that is several orders
of magnitude greater than the present age of the universe. Inflaton decay
is related to an interaction which, together with a spontaneous breakdown
of CP invariance at a cosmological energy scale, can give rise to a
neutrino-antineutrino asymmetry just prior to the time of electroweak
symmetry breaking.}
\section{Introduction}
The purpose of this paper is to present the results of calculations
which support two new ideas concerning matter in the universe:
\begin{itemize}
\item[(1)] A large part of dark matter in the present universe can be 
composed of massive, cold inflatons, the scalar quantum of the field
whose vacuum energy initiated a period of inflation \cite{ref1,ref2,ref3}.
\item[(2)] There is a definite possibility that the inflatons are not
absolutely stable; they can decay in a specific way, with a possible lifetime
estimated here to be several orders of magnitude greater than the
present age of the universe. 
\end{itemize}
The above results are based upon detailed calculations within a specific
dynamical model \cite{ref4}. These calculations show, first, that 
radiative corrections to the tree-level potential for the scalar
field, $\phi$, can set up inflation \cite{ref4}. The radiative corrections
are calculated via the renormalization-group equations, using the
potential for $\phi$ and quantum interactions for definite additional
fields: a massive, neutral lepton $L$, and a massive, spin-zero boson $b$.
(The primordial fields are electrically neutral.) The calculated 
radiatively-corrected potential has a maximum at an energy scale near to the Planck
scale $M_P=1.2\times 10^{19}\GeV$, and a minimum below this scale,
at $\phi = \phi_c$. We consider that inflation occurs while the scalar
field rolls down from this maximum toward the minimum. (Much can occur at the
maximum.)\par
In obtaining the above results (1,2) we shall use explicit, calculated 
examples of the radiatively-corrected potential, $V_c(\phi)=\lambda(\phi)\phi^4$;
these potentials are shown in Fig.~1 and Fig.~2. We briefly recapitulate
the method of calculation\cite{ref4}. The potential and quantum interactions
are given by\footnote{For pseudoscalar $b$, $\ol{\psi}_L\psi_Lb\to 
i\ol{\psi}_L\gamma_5\psi_Lb$. The results in Figs.~1,2 are unchanged}.
\be
V(\phi,b)=\lambda\phi^4 + \lambda_b\phi^2b^2 + \tilde{\lambda}_bb^4 + 
g\ol{\psi}_L\psi_L\phi + g_b\ol{\psi}_L\psi_L b
\ee
All coupling parameters are dimensionless. The equations giving the
development of the couplings with the energy scale $\phi$ are
\ba
16\pi^2\frac{d\lambda(\phi)}{dt} &=& 72\lambda^2 + 2\lambda_b^2 - 2g^4 =
 \beta_\lambda(\phi)\nn\\
16\pi^2 \frac{d\lambda_b(\phi)}{dt} &=& \lambda_b(24\lambda + 16\lambda_b
 + 24\tilde{\lambda}_b) - 4g^2g_b^2 = \beta_{\lambda_b}(\phi)\nn\\
16\pi^2 \frac{d\tilde{\lambda}_b(\phi)}{dt} &=& 
 72\tilde{\lambda}_b^2 + 2\lambda_b^2 - 2g_b^4 = \beta_{\tilde{\lambda}_b}(\phi)\\
16\pi^2 \frac{dg(\phi)}{dt} &=& 5g (g^2 + g_b^2) = \beta_g(\phi)\nn\\
16\pi^2 \frac{dg_b(\phi)}{dt} &=& 5g_b (g_b^2 + g^2) = \beta_{g_b}(\phi)\nn
\label{beta}
\ea
The extremum conditions \cite{ref4,ref5} are imposed at the scale 
$\phi=\phi_i\simlt M_P$, at which the computations are begun,
\ba
\beta_\lambda(\phi_i)=0\nn\\
\beta_{\lambda_b}(\phi_i)=0
\ea
For given values of $\lambda(\phi_i)$ and $\lambda_b(\phi_i)$ (with
\footnote{Smaller $\tilde{\lambda}_b(\neq \lambda_b)$ move the calculated
maximum and minimum further away from $\phi_i$. $\tilde{\lambda}_b$ can
be made much smaller, and the effect compensated by increasing $\lambda_b$
slightly.}
$\tilde{\lambda}_b(\phi_i)=\lambda_b(\phi_i)$), these conditions
\underline{determine} $g(\phi_i)$ and $g_b(\phi_i)$. The chosen values
for $\lambda(\phi_i)$, $\lambda_b(\phi_i)$ were determined \cite{ref4},
in order of magnitude, by conditions during the period of inflation,
i.~e.~``slow roll-down'' and ``sufficient expansion''\cite{ref6}. In
this paper, we deal with physical processes which occur at times
after $\phi$ has fallen through the minimum of the potential. The
renormalization group equations (2) are solved numerically. The potential
shown in Fig.~1 has $\phi_i = (4\times 10^{-2})M_P$, 
$\lambda(\phi_i)=10^{-13}$ and $\lambda_b(\phi_i)=10^{-4}$. The calculated
position of the maximum is $\phi=\phi_m=M_P$, and the calculated position
of the minimum is $\phi=\phi_c=10^{-3}M_P$. The calculated initial
values for the coupling parameters are $g(\phi_i)=10^{-2}$, and
$g_b(\phi_i)=3.2\times 10^{-2}$. The potential shown in Fig.~2 has 
$\phi_i=10^{-1}M_P$, $\lambda(\phi_i)=4.3\times 10^{-14}$, and 
$\lambda_b(\phi_i)=10^{-4}$. The maximum is at $\phi_m=M_P$; the minimum
is at $\phi_c=10^{-2}M_P$. The initial couplings are $g(\phi_i)=10^{-2}$
and $g_b(\phi_i)=3.2\times 10^{-2}$. In addition, the mass of the inflaton is calculated
from $d^2V_c(\phi)/d\phi^2|_{\phi=\phi_c}=m_\phi^2$; this is
$(6\times10^9\GeV)^2$ for Fig.~1 and $(5\times 10^{10}\GeV)^2$ for Fig.~2.
\section{The matter energy density of inflatons}
The discussion which follows on the basis of these potentials is largely
phenomenological, as must be the case for a discussion of interaction
and decay processes which involve hypothetical massive particles in the
early universe. The numerical work is meant to be indicative of a general
picture for the development in time of the matter (and radiation) energy
densities. Consider first the (squared) fluctuations $(\delta\phi)^2$ of
the inflaton field in the vicinity of the minimum. We consider that there
is a matter energy density $\rho_M^\phi$, given by the inflaton (squared)
mass,
\be
\rho_M^\phi(t_c)\cong\frac{1}{2}m_\phi^2(\delta\phi)^2
\ee
The fluctuations can develop semiclassically\cite{ref7} to a maximum value
\cite{ref8,ref9}, which we parametrize in two alterative forms
\footnote{The form (5a) for ``scaled'' fluctuations is well-known in
high energy particle collisions. Under the name \cite{ref10,ref11}
``KNO scaling'', it provides an approximate representation, in a certain
energy interval, for the fluctuations in the number of produced particles
about the average.($(\delta\phi)^2$ denotes a time-average.)} 
\begin{subeqns}
(\delta\phi)^2 \cong \epsilon\phi_c^2\\
(\delta\phi)^2 \cong \epsilon\phi_c\phi_i
\end{subeqns}
where $\epsilon$ is a small parameter. As we shall discuss below, independent
estimates \cite{ref8,ref9} give values which lie in the interval
\be
10^{-7}\le\epsilon\le10^{-5}
\ee
Note that $\epsilon$ is approximately given by $\sqrt{\lambda}$.
We assume that the squared mass $m_\phi^2$, and the maximal squared 
fluctuations $(\delta\phi)^2$, together provide a measure of the initial matter
energy density. This at about a time $t_c\cong (1/\phi_c)\cong 0.6\times 10^{-40}
\mathrm{sec}$, which we use as the ``starting'' time (after inflation) in
our calculation of the energy density\footnote{There is a time interval between
$t_c$ and the full development of the fluctuations (note ref.~9). Taking
$t_c$ as the approximate ``starting'' time for evolution of the full
energy density gives a lower bound on the later densities, for a given
initial density fixed by the specific choice of parameter $\epsilon$. In 
other words, increasing the effective $t_c$ to $t_c'$ allows
for a decrease in $\epsilon$ as $(t_c/t_c')^2$; this means that only
(possibly much) smaller fluctuations are necessary, for the
results in this paper. For example, consider $t_c'\sim 10^{-36}$ sec. With
$(\delta\phi)^2\sim m_\phi^2 \sim 10^{-13}\phi_c^2$ one has
$\rho_M^\phi(t_c')\sim 2.5\times 10^{38} \GeV^4$, and $\rho_M^\phi(t_0)\sim
5\times 10^{-47}\GeV^4$ evolves from $t_c'$, as in eq.~(9) below.}.
With reference to the calculated potential in Fig.~1 ($\phi_c=10^{-3}M_P$, 
$m_\phi^2\cong (5 \times 10^{-10}M_P)^2$), 
and using eq.~(5a) with $\epsilon\cong 10^{-5}$, gives a matter energy density
\footnote{Note that $\lambda(\delta\phi)^2 < m_\phi^2$} for massive,
cold inflatons at $t_c$ of
\be
\rho_M^\phi(t_c)\cong
\frac{1}{2}m_\phi^2(\delta\phi)^2\cong\frac{1}{2}(5\times 10^{-10}M_P)^2 (10^{-11}M_P^2)
\cong 2.5\times 10^{46}\GeV^4
\ee
Use of eq.~(5b), with $\epsilon=10^{-7}$, gives $\rho_M^\phi(t_c)\sim 10^{46}\GeV^4$.
We assume the inflaton to be absolutely stable, or, as we shall show below, stable
up to time scales much greater than the present age of the universe; for the latter
we use in our calculations the approximate value $t_0\cong 4\times 10^{17}\sec$.
Note that the masses of the fermion $L$ and the boson $b$, to which $\phi$ is
coupled by the interactions in eq.~(1), are approximately given by
\ba
m_L\cong g(\phi_c)\phi_c\sim (10^{-2}) (10^{16} \GeV) \cong 10^{14}\GeV\nn\\
m_b\cong \sqrt{\lambda_b(\phi_c)}\phi_c\sim \sqrt{10^{-4}} (10^{16} \GeV) \cong 10^{14}\GeV\nn\\
\ea
Both masses are much greater than $m_\phi$. We estimate the evolution of the matter
energy density in eq.~(7) from $t_c$ to $t_0$, using three distinct intervals. First,
a relatively brief time interval up to $t_H\cong 0.7\times 10^{-34} \sec$, in
which we assume that the universe is matter dominated. Here $t_H$ is a time
(calculated in section 3 below), near which a massive ($m_H\cong 10^{14}\GeV$), 
spin-zero boson $H$, decays into two photons, giving rise to a radiation energy
density of about $m_H^4\cong 10^{56}\GeV^4$. We then consider a second interval
in which radiation dominates. This interval lasts until the usual time\cite{ref6}
of transition back to matter dominance, for which we use in our calculations 
$t_M\cong 10^{11}$ sec. The third interval is then from $t_M$ to $t_0$.
%
% pg 7 of manuscript
%
The matter energy density of inflatons $\rho_M^\phi$, is proportional to $R^{-3}(t)$;
the scale factor $R(t)$ develops as $t^{2/3}$ when matter dominates, and as $t^{1/2}$ when
radiation dominates. Using eq.~(7), the result for the matter energy density for
massive, cold inflatons in the present universe is then
\ba
\rho_M^\phi(t_0) &\cong& \rho_M^\phi(t_c)\left(\frac{0.6\times 10^{-40}}{0.7\times 10^{-34}}\right)^2
\left(\frac{0.7\times 10^{-34}}{10^{11}}\right)^\frac{3}{2} \left( \frac{10^{11}}{4\times 10^{17}}\right)^2\nn\\
&\cong& (2.5\times 10^{46} \GeV^4)(0.75\times 10^{-93}) \cong 2\times 10^{-47}\GeV^4
\ea
For the case of eq.~(5b) with $\rho_M^\phi(t_c)\cong 10^{46} \GeV^4$, $\rho_M^\phi(t_0)\cong 10^{-47}\GeV^4$.
These numbers are rather close to the closure energy density of about $(2-3.5)\times 10^{-47}\GeV^4$.
A value $\rho_M^\phi(t_0)\cong 1.5 \times 10^{-47}\GeV^4$ is calculated from the potential
in Fig.~2 ($\phi_c = 10^{-2}M_P$, $m_\phi^2 \cong (4.25\times 10^{-9}M_P)^2$),
using eq.~(5a) with $\epsilon \cong 10^{-7}$. Regard\-ing the energy scale $\phi_c\cong 10^{16}\GeV$
(equivalently, the time$^{{\mathrm F}4}$ $t_c\cong (1/\phi_c)$) at the end of inflation, 
it is noteworthy
that a comparable scale has been obtained in a recent analysis \cite{ref12} made to determine an upper-limit 
value for the Hubble parameter $H(t)$, at the end of inflation: $(H_{\mathrm{end}})_{\mathrm{max}} \cong 4\times 10^{-6} M_P$
(of the order of $3\times \phi_c^2/M_P)$.\par
The result for $\rho_M^\phi$ depends upon the size of the 
fluctuations$^{\mathrm{F}4}$ from eqs.~(5,6). Therefore,
we summarize the elements in recent calculations \cite{ref8,ref9} which 
estimate\footnote{In ref.~8, the estimate has a large element of assumptions concerning initial conditions.
In ref.~9, the detailed numerical simulation of the evolving system also rests upon a number
of assumptions about initial conditions.} the parameter $\epsilon$. In ref.~8, $\epsilon$
is the product of two factors. One factor is a squared ``initial'' amplitude for oscillation,
assumed to be $\sim (10^{-1}M_P)^2$, around an ``initial'' scale for $\phi$, which is assumed
to be $\phi\cong M_P$. The second factor in effect reduces the final squared fluctuation
because of the finite time for build-up; this factor depends explicitly upon $\lambda$, calculated
as $\sim(\ln(1/\lambda))^{-2} \cong 10^{-3}$. Thus, the value of $\epsilon$ is $\sim (10^{-2})(10^{-3})
= 10^{-5}$ (in this case, the coefficient of the assumed ``initial'' value of $\phi\cong M_P$
at the end of inflation). In ref.~9, two factors are again involved; a smaller value
of $\epsilon\cong 10^{-7}$ is obtained for fluctuations in non-zero wave-number modes, 
largely because of a longer time interval for build up.
The authors \cite{ref9} have used some initial-condition fluctuations about which 
they only state: ``the initial fluctuations are of the correct order of magnitude'',
and that the parameter $\lambda$ ``regulates the magnitude of initial quantum fluctuations
in non-zero (wave-number) modes relative to the magnitude of the zero mode''.  The
latter is implicitly assumed to be close to $\phi=M_P$. The essential physical point is that
values of $\epsilon$ in the range given in eq.~(6) can represent a 
maximum parameter$^{\mathrm{F}4}$ in the estimate of the fluctuation amplitude  $|\delta\phi|$. 
On the other hand, our calculated value
of $\phi_c\cong 10^{-3}M_P$, the classical field variable after inflation, is not as close to $M_P$
as the values assumed and used, without calculation, in the papers \cite{ref8,ref9} whose results
we have summarized. From eqs. (5a,b), this is relevant for the size of 
$(\delta\phi)^2$.
\section{Decay of the inflaton}
A neutral lepton $L$ (neutrino-like) and its antiparticle $\ol{L}$, play a role in virtual
intermediate states in the dynamical model\cite{ref4} for calculating the radiatively-corrected,
tree-level potential for the inflaton field (Fig.~1 and Fig.~2). The lepton $L$ is massive,
from eq.~(8) $m_L\cong 10^{14}\GeV$, as determined by the calculated values for $\phi_c$ and
for $g(\phi_c)$. \par
There is an interesting way in which massive $L$ and $\ol{L}$, essentially at rest, can
disappear: this is via a decay process caused by a minuscule mixing with a light neutrino,
assuming that the latter has mass, say $m_{\nu_\tau}\sim 1.8\;\mathrm{eV}$. This possible
process has special interest because it can lead to the eventual decay of the inflaton. In
section 4, we show that the decay of $L$ and $\ol{L}$ also has possible relevance for the
generation of a matter-antimatter asymmetry, specifically a neutrino-antineutrino-asymmetry
which can occur at a time just somewhat prior to the electroweak symmetry-breaking time,
$\sim 10^{-12}$ sec. Consider the decay
\be
L(\ol{L}) \to \nu_\tau(\ol{\nu}_\tau) H
\ee
We assume that the decay occurs because of a mixing interaction of the specific strength\cite{ref13}
\footnote{The hypothesis of this kind of coupling between light neutrino mass states\cite{ref13},
is being tested by the CHOOZ reactor experiment\cite{ref14}, starting with $\ol{\nu}_e$.
It can accomodate $m_{\nu_\mu}\simlt 0.03$ eV (for $m_{\nu_\tau}\sim 1$ eV) and
sizable $\nu_e-\nu_\mu$ mixing. (Note that the alternative $m_{\nu_\mu}\sim m_{\nu_\tau}$
with large mixing, can be accomodated.)} 
$$
\left(\sqrt{\frac{m_{\nu_\tau}}{m_L}}\right) 
\left( \ol{\psi}_{\nu_\tau} \psi_L H + \ol{\psi}_L \psi_{\nu_\tau} H\right)
$$
The spin-zero boson $H$ is Higgs-like (or $\sigma$-like) in that its coupling to fermions
contains a mass factor (in effect $\sqrt{m_{\nu_\tau}m_L}$, relative to a scale $F_H\sim m_H\sim m_L$).
In general, with a coupling which mixes neutrino mass states, it is a distinct boson with 
a mass $m_H=rm_L$, $r\simlt 1$, i.~e.~of the order of $10^{14}\GeV$. The decay width is
\be
\Gamma(L(\ol{L})\to\nu_\tau(\ol{\nu}_\tau)H) = \frac{(1-r^2)^2}{8\pi}m_{\nu_\tau} \sim 10^{14}\sec^{-1}
\ee
where we have used $m_{\nu_\tau} \cong 1.8$ eV, and $(1-r^2)\sim 1$. Because of the specific strength
of the coupling, this decay width is \underline{independent} of $m_L$, depending instead upon
the assumed non-zero value of $m_{\nu_\tau}$. The lifetime is $\tau_L\cong 10^{-14}$ sec, 
a number that is notable because it appears naturally close to the time of electroweak symmetry 
breaking, $\sim 10^{-12}$ sec.
\par
Now, \underline{independently} of whether or not there is an ``initial'' energy density of $L$
and $\ol{L}$ at $t_c$, the inflaton will, in general, decay via a virtual intermediate state of an $L\ol{L}$
pair: $\phi\to\nu_\tau \ol{\nu}_\tau$. The Feynman graph is shown in Fig.~3. Calculation of
the decay width gives,
$$
\Gamma(\phi\to \nu_\tau\ol{\nu}_\tau)=g^2\left(\frac{m_{\nu_\tau}}{m_L}\right)^2\left(\frac{m_\phi}{8\pi}\right)
|{\cal M}|^2 \cong\left(\frac{m_{\nu_\tau}}{\phi_c}\right)^2 \left(\frac{m_\phi}{8\pi}\right)|{\cal M}|^2
$$
where 
\be
|{\cal M}| \cong \frac{1}{16\pi^2}\ln\left(\frac{\Lambda^2+m_L^2}{m_L^2}\right) \sim 10^{-2}
\ee
In eq.~(12), we have used $m_L\cong g \phi_c$; thus $\Gamma_\phi$ is \underline{independent} of g.
$\Lambda$ which is taken as $\sim 2m_L$ for an approximate estimate, is a cut-off on the
intermediate-state momentum integral. Because of the minuscule number $(m_{\nu_\tau}/\phi_c)^2\cong
3.2\times 10^{-50}$, the lifetime for inflaton decay is,
\ba
\tau_\phi = \Gamma^{-1}_\phi &\cong& \left(\frac{\phi_c}{m_{\nu_\tau}}\right)^2
\left(3.3\times 10^{-29}\;\mathrm{sec}\right)\nn\\
&\cong& 10^{21}\;\mathrm{sec}
\ea
Thus, we have used the calculated value of $\phi_c\cong 10^{16}\GeV$ for the potential in Fig.~1,
to estimate a possible inflaton lifetime about three orders of magnitude greater\footnote{
For spin-zero $\phi$, one can make the lifetime much larger
by applying $(1+b\gamma_5)$ to $\psi_{\nu_\tau}$ in the mixing interaction
(after eq.~(10)), and letting $b$ approach unity. This can be a reason for the
projection, implying parity nonconservation.} than the present age of
the universe. Under the assumption that $\phi$ decays in this way, one may make an anthropic
type of argument to the effect that $m_{\nu_\tau}$ should not be greater than about 60 eV, since
$\tau_\phi>t_0$ must hold.
Using $t_0\sim 10^{18}$ sec, eq.~(13) gives $(10^{21})(1.8/m_{\nu_\tau})^2 > 10^{18}$, 
and therefore $m_{\nu_\tau}<60$ eV. It is worth noting the closeness of this number
to the Cowsik-McClelland bound\cite{ref15} for the mass of a stable, light neutrino species,
$m_\nu<91.5$ eV, (obtained by taking the reduced Hubble parameter $h(t_0)<1$).\par
Up to this point, we have not invoked any assumption concerning whether or not there is a
significant ``inital'' energy density (at $t\simgt t_c$) for $L$ and $\ol{L}$ (and/or for
$b$ quanta). We discuss this question in section 4, which concerns a neutrino-antineutrino
asymmetry.
The $H$ boson invoked for the decay in eq.~(10), exhibits a coupling to a light neutral
lepton; therefore it could naturally be considered to decay promptly into two photons 
(presumably via a characteristic triangle-graph anomaly\cite{ref16,ref17} involving 
charged particles). To calculate the lifetime, we use the interaction $(e^2/f_H)(\epsilon_{\mu\nu\sigma\rho}
F_1^{\mu\nu}F_2^{\sigma\rho})H$, where $F_{1,2}^{\mu\nu}$ are the field tensors for the two photons,
$e^2$ is the squared electric charge, and the decay constant is a parameter, which we take
as $f_H\sim F_H \sim m_H \sim 10^{14}\GeV$. The lifetime for the $H$ quantum is then
\be
\tau_H\cong\left(\Gamma(H\to2\gamma)\right)^{-1} = \left(\left(\frac{e^2}{4\pi}\right)^2
\left(\frac{\pi}{2}m_H\right)\right)^{-1} \cong0.7\times 10^{-34}\;\mathrm{sec}
\ee
This is the time $t_H$ used in eq.~(9) for the end of the first time interval in which
matter dominates, and the beginning of the second time interval in which radiation
dominates. For example, considering only photons, an energy density $\rho_M^H \sim m_H^4
\sim 10^{56}\GeV^4$, assumed to be present\footnote{As a means of initiating radiation,
after inflation, this is more direct than invoking, under the assumption of some
grand unification, a large number of hypothetical, massive particles, produced in
hypothetical non-thermal conditions, and disappearing into radiation in such conditions,
via a series hypothetical decays. In any case, the maximal ``thermal'' condition defined
\cite{ref9} by an effective temperature of $T_{\mathrm{eff}}=\sqrt{12(\delta\phi)^2}$ is $\sim 10^{14}\GeV$
according to eq.~(7). Note that $m_H\cong m_b \cong m_L$, with $H\to 2\gamma$, produces the
same energy scale for radiation, since $m_b\cong m_L\cong 10^{14}\GeV$ is
\underline{calculated} in eq.~(8).}
at about $t_H\cong 10^{-34}$ sec, and the decay $H\to 2\gamma$, leads to a radiation
energy density which evolves to a present-day value of
\ba
\rho_\gamma(t_0)&\sim& \rho_M^H(t_H)\left(\frac{10^{-34}}{10^{11}}\right)^2
\left(\frac{10^{11}}{4\times10^{17}}\right)^{8/3}\nn\\
&\cong& (10^{56}\GeV^4)(0.25\times10^{-107}) \cong 0.25\times 10^{-51}\GeV^4
\ea
A relevant point about the above $\rho_M^H(t_H)$ (as well as about other possible$^{\mathrm{F}10}$
(cold) matter energy densities such as $\rho_M^{L,\ol{L}}(t_H)$, $\rho_M^b(t_H)$),
is that the values lie well below the maximum total energy density at the end of inflation
which has been estimated\cite{ref12} from the upper-limit value for the Hubble parameter 
at the end of inflation.
This energy density is\cite{ref12} $\sim \frac{1}{8}((H_{\mathrm{end}})_{\mathrm {max}} M_P)^2\sim
4\times 10^{64}\GeV^4$. Another point is that massive $\phi$ quanta, and possible $L$, $\ol{L}$
quanta, have virtually no interaction contact with surrounding thermal conditions in the model
of eq.~(1), from which the potential in Fig.~1 is calculated.
\section{CP noninvariance and a neutrino-antineutrino asymmetry}
A neutral, spin-zero boson $b$ plays an essential role in the dynamical model\cite{ref4} for
calculating the radiatively-corrected, tree-level potential for the inflaton field.
We consider $b$ to be a pseudoscalar (see appendix); from eq.~(8), it is massive, $m_b\cong 10^{14}\GeV$,
as determined by $\lambda_b(\phi_c)$ and $\phi_c$. Such a pseudoscalar boson can play 
an important role in the early universe, where it can be the cause of the CP noninvariance
that is necessary to arrive at a matter-antimatter asymmetry. CP noninvariance occurs
spontaneously\cite{ref18} if the $b$-field acquires a non-zero vacuum expectation value, $F_b$.
In the following discussion, we shall assume that CP violation occurs, at least
partly, through a spontaneous breakdown at a cosmological energy scale, (we have
previously estimated\footnote{This estimate comes from consideration of the mixing
of CP odd and even states in the $(K^0_L-K^0_S)$-system, where this mixing is viewed
as a remnant of the breakdown of discrete symmetries at a cosmological energy scale.
(In ref.~19, note eq.~(13) with $M_P$ replaced by $m_L$. The small ratio of squared
energy scales appears as $(F_b^2/M_P^2)$ because $m_L$ was assumed to be close
to $M_P$, which is not the case in the present work. The primary time interval
used in ref.~19 is set by the Hubble parameter $(H(m_L))^{-1}$ i.~e.~by
the energy scale $m_L$.) Note that with $F_b\ll \phi_c$, corrections
to the masses estimated in the present paper are small. $\Gamma(b\to\phi\phi) 
\propto F_b^2\cong (3.5\times 10^{-23}\;\mathrm{sec})^{-1}$. If $m_b\simgt 2.2 m_L$,
$\Gamma(b\to L\ol{L}) \simgt (1.5\times 10^{-33}\;\mathrm{sec})^{-1}$ is much larger.
This can be the origin of $L,\ol{L}$. 
For example, with $\rho_M^b(t\sim 1/F_b\sim 2\times 10^{-35}\sec)\sim 10^{46}\GeV^4$.}\cite{ref19} this to be at
about $F_b\sim 4\times 10^{10}\GeV$). Since we have shown that the decay processes
$L(\ol{L})\to \nu_\tau (\ol{\nu}_\tau)H$ occur at a time just some\-what prior to the
time of electroweak symmetry breaking, they represent natural processes for generating
an asymmetry in the number densities for neutrinos and antineutrinos. Such an asymmetry can
result in a 
baryon-antibaryon asymmetry via processes which occur in the course of
electroweak symmetry breaking \cite{ref20,ref21}. \par
We have examined a novel mechanism for obtaining the asymmetry, one which relies explicitly
on the spontaneous breakdown of CP invariance at a cosmological energy scale. The description
of the process given here is intended only as a semi-quantitative \underline{example}, one
which is based upon the particle content and the interactions that occur in the dynamical
model which produced the radiatively-corrected, tree-level potential in Fig.~1, and
which allows for eventual decay of the massive, cold quanta of the inflaton field. The
unusual aspect of the decay process is illustrated by the Feynman graphs in Fig.~4,a,b. 
In (a) the decay process occurs for an isolated $L$; in (b) the process
is different, the decay occurs for an $L$ which we consider to be in a quasi-bound,
metastable state with an $\ol{L}$. The
quasi-bound state involves the exchange of $b$-quanta i.~e.~there is an ``initial-state'' interaction
which gives rise to complex ``scattering'' amplitudes that multiply the decay vertex. There are, 
in fact, two different such amplitudes, corresponding to $\gamma_5$ and $1$ at the upper
$b$-vertex in Fig.~4b.
The spontaneous CP violation generates\cite{ref18} a CP-violating, quantum interaction like
$(L\ol{L})b$ which has a scalar interaction character, starting from the CP-invariant,
pseudoscalar interaction $ig_b(\ol{L}\gamma_5 L)b$. The scalar interaction is essential
to have a non-zero lower vertex in Fig.~4b. Together with the decay vertex, of
the general form $(a(\ol{L}\nu_\tau + \ol{\nu}_\tau L)+b(\ol{L}\gamma_5\nu_\tau -
\ol{\nu}_\tau\gamma_5 L))H$ (this does \underline{not} violate CP), all of the
elements are present that are necessary for obtaining a CP-violating difference between
the partial rates for the particle and the antiparticle decay processes, as they occur
in the quasi-bound configurations in Fig.~4b, $L(+\ol{L})\to \nu_\tau H(+\ol{L})$,
$\ol{L}(+L)\to \ol{\nu}_\tau H(+L)$. The asymmetry in neutrino number density is proportional
to the difference in the branching fractions for the process with
particle and the process with antiparticle. Since there are initially equal numbers of
 $L$ and $\ol{L}$, the asymmetry vanishes if the branching fractions are unity. There
must be at least two processes by which $L$ and $\ol{L}$ ``disappear''. We give the result
of an estimate for the number-density asymmetry,
\ba
|A(n_{\nu_\tau}-n_{\ol{\nu}_\tau})|&\cong& \left\{\frac{\Gamma(L(+\ol{L})\to \nu_\tau H(+\ol{L}))}
{\Gamma(L\to\nu_\tau H)} \right\}\nn\\
&\times&\left\{\frac{|\Gamma(L(+\ol{L})\to \nu_\tau H(+\ol{L})) -
 \Gamma(\ol{L}(+L)\to \ol{\nu}_\tau H(+L))|}{\Gamma(L(+\ol{L})\to \nu_\tau H(+\ol{L}))}\right\}\nn\\
&\sim& ({\mathrm{b.~r.~}}) \left\{\frac{g_b^2\langle v^2\rangle^{3/2}}{4\pi}\right\}
\ea
The first factor in eq.~(16) is essentially the  branching fraction for
the decay process to occur from the quasi-bound state configuration; $\langle v^2\rangle^{1/2}$
represents a characteristic \underline{internal} velocity for $L(\ol{L})$ in
the quasi-bound state configuration; and $g_b^2\cong 10^{-3}$ has been calculated
(for the potential in Fig.~1). The factor in curly brackets in eq.~(16) is the result of the
``initial-state'' interaction. As an example, with (b.~r.~)$\sim 0.1$ and $\langle v^2\rangle^{1/2}
\simlt 1$, the number $\simlt 10^{-5}$ from eq.~(16) results in a small overall neutrino-antineutrino
asymmetry. This is because the overall asymmetry must depend upon the number density of massive
$L(\ol{L})$ that are present at the decay time, relative to the number density of quanta
in the radiation field at this time; this small ratio dilutes the asymmetry. Also, the
CP-violating, scalar interaction is probably depressed
by a factor like $(F_b/m_L)$ (see appendix). Nevertheless, the relatively high energy
involved in the neutrino-antineutrino asymmetry might allow an enhanced quark-antiquark
number asymmetry at lower energy, essentially compensating the above-mentioned dilution
factor.\par
The appearance of at least some part of a quark-antiquark asymmetry driven by an asymmetry between 
neutrinos and antineutrinos of relatively high
energy, at the time of electroweak symmetry breaking, suggests an unusual possibility.
The appearance of some baryons formed from quarks may then occur at a time 
after the time of photon decoupling, thus providing
an impulse to structure formation.
\section{Vacuum energy after inflation}
There is a particular vacuum energy density at $\phi_c$ ( at time $\sim t_c$), 
$V_c(\phi_c)=\lambda(\phi_c)\phi_c^4
\sim -2\times 10^{50}\GeV^4$. The magnitude is only similar to the matter energy densities
of massive, cold quanta which are present in the first time interval after inflation
that we have considered in this paper, $t_c\le t \le t_H$. Consider a time-variation
of $|V_c(\phi_c)|$. If $|V_c(\phi_c)|$ were to fall like radiation, i.~e.~proportional
to $R^{-4}(t)$, it would fall by about four orders of magnitude relative to the matter
energy densities in the first interval. The assumption of matter dominance due to 
massive, cold $\phi$, and $H$, quanta in the first time interval, prior to $H\to 2\gamma$,
is made. If this fall were to continue, the magnitude of the vacuum energy density comes
to minuscule values, $\sim 10^{-34}\GeV^4$ at the time of nucleosynthesis $\sim 1$ sec,
and $\sim 10^{-73}\GeV^4$ at $t_0$. If $|V_c(\phi_c)|$ were to
fall like a matter energy density from about $10^{-34}$ sec to $\sim 1$ sec, a still relatively
small value of $\sim 10^{-17}\GeV^4$ occurs at the nucleosynthesis time. If continued to
$t_0$, $\sim 2 \times 10^{-47}\GeV^4$ occurs; this is in itself interesting, being in
magnitude similar to $\rho_M(t_0)$ in eq.~(9).\par
A speculation involves the possible transfer of this energy to binding in an
$L\ol{L}$ ``condensate'', which decays prior to the time of electroweak symmetry breaking,
as discussed in section 4. An estimate indicates that this binding
could arise from the particular force which involves the trilinear
couplings proportional to $(4\lambda\phi_c)$ in eq.~(A8) of the
appendix.\par
Alternatively, of course, the vacuum energy density at the position of the potential
minimum can be adjusted to zero \cite{ref4}. Equivalently, this negative energy density
can be cancelled in a brief time by another positive density, such as that
associated with the related field $b$ (see appendix). Note that $\lambda_b\phi_c^2 F_b^2$
is $\sim 2\times10^{49}\GeV^4$, similar to $|V_c(\phi_c)|$ above.\par
Of the above possibilities, perhaps participation in $L\ol{L}$ binding is particularly
intriguing, because $|V_c(\phi_c)|$ is in effect eliminated.\par
The discussion in this section leaves open the possibility that a positive
residual vacuum energy density which decreases with increasing time can
contribute to a relevant effective cosmological constant in the present
universe. It is perhaps noteworthy that a vacuum energy density of about
$+2\times 10^{50}\GeV^4$ at $\sim 10^{-34}\;\sec$, which decreases as matter
to the time of $(L-\ol{L})$ decay $\sim 10^{-14}\; \sec$ and as radiation
thereafter, attains a value of $\sim 5\times 10^{-48}\GeV^4$ at present.
\section{Conclusions}
Models for the period in the universe following a hypothetical period
of inflation, which are based upon particle physics, tend to invoke somewhat
arbitrary interaction forms and coupling strengths. In addition, there are
a number of constraints imposed upon the dynamics in order to obtain some
desired effects. Practically all of the hypotheses which are invoked cannot be tested
(note footnote F9).\par
There are also primary questions: just how does the dynamics of a scalar field 
set up inflation, and what is the further role of this scalar field in the evolution
of the universe?\par
In this paper, we have motivated our considerations by explicit, detailed calculation
of the radiatively-corrected, tree-level potential which sets up inflation\cite{ref4}.
This potential is exhibited in Fig.~1, with its calculated maximum where inflation
begins and calculated minimum where inflation ends. The calculation is carried out
within a dynamical model with only a few essential fields, that is in addition
to the inflaton field $\phi$, a massive, neutral lepton $L$ and anti-lepton $\ol{L}$,
and a massive, spin-zero boson $b$. The masses are calculated approximately from
the potential and coupling parameters. A phenomenological consideration of the
evolution of the matter energy density indicates the possibility that quanta of the
inflaton field constitute a significant part of cold, dark matter in the present universe.
When we consider a possible characteristic of the mixing between $L$ and a light
neutral lepton like $\nu_\tau$, assumed to have a mass, the possibility arises of
the eventual decay of the inflaton.
This occurs via virtual $L\ol{L}$: $\phi\to\nu_\tau\ol{\nu}_\tau$, with a calculable
lifetime which depends upon the ratio $(\phi_c/m_{\nu_\tau})^2$. There are tests
of these ideas. One involves comparision of the dark-matter energy density that
is actually required by measurements, with that obtained for inflatons in improved
calculations. A second involves testing the hypothesis that the mixing of
neutrino mass states occurs via a coupling proportional to the square root of the
mass ratio. The neutrino oscillation experiments can eventually do
this, within the system of light neutrinos$^{\mathrm{F}7}$. There, the hypothetical
pattern involves one sizable mixing, either between $\nu_e$ and $\nu_\mu$, or
between $\nu_\mu$ and $\nu_\tau$; cancellations \cite{ref13} cause the other mixings
to be much smaller. The determination
of the mass of the heaviest of these, presumably $m_{\nu_\tau}$, is 
relevant for the question of hot, dark matter, as well as to inflaton decay.
A third direction involves examination of the possibility that CP invariance
breaks down spontaneously at a cosmological energy scale\cite{ref19}. If this
is the only source of CP noninvariance (i.~e.~the quark mixing matrix is real),
then the mixing of CP odd and even states in the $(K_L^0 - K_S^0)$ system must
be a remnant\cite{ref19} of this high-scale symmetry breaking. CP violation
primarily involving leptons enters the hadronic sector much depressed, via leptonic
intermediate states. The analogue for neutral leptons of the mixing in the
neutral kaon system is $\ol{\nu}_\mu \nu_e\to \nu_\mu\ol{\nu}_e$.\par
The above ideas, which we have illustrated quantitatively in this paper, provide
some new possibilities, perhaps worthy of further study.
\section*{Appendix: $\phi$, $b$, and $L$ in a primordial chiral symmetry, broken
spontaneously and explicitly}
\renewcommand{\theequation}{A\arabic{equation}}
\setcounter{equation}{0}
There is a ``toy'' Lagrangian model which contains a number of aspects of the
dynamics that we have discussed in this paper. The Lagrangian illustrates the
possibility of a primordial chiral symmetry involving
a neutral scalar $\phi$, a neutral pseudoscalar $b$, and a neutral lepton $L$. The
symmetry is broken both spontaneously and explicitly. The model is patterned after
the $\sigma$-model of pion physics\cite{ref22,ref23}. The Lagrangian density is\cite{ref24}
\ba
{\cal L} &=& \frac{\i}{2} \left(\ol{L}\gamma_\mu(\partial^\mu L) - (\partial^\mu \ol{L})
\gamma_\mu L\right) + \frac{1}{2}\left((\partial_\mu\phi)^2 + (\partial_\mu b)^2\right)\nn\\
&-&\frac{\mu^2}{2}\left(\phi^2+b^2\right) - g \ol{L}L\phi - \i g \ol{L}\gamma_5 L b \nn\\
&-&\lambda \left(\phi^2+b^2\right)^2
\ea
${\cal L}$ is invariant under the chiral transformation (with infinitisimal, constant $\beta$)
\ba
L\to L - \frac{\i\beta}{2}\gamma_5 L && \ol{L}\to \ol{L} - \frac{\i\beta}{2}\ol{L}\gamma_5\nn\\
\phi\to\phi-\beta b && b\to b + \beta \phi
\ea
As a consequence $\partial^\mu A_\mu(x) = 0$, where a conserved axial vector current is given by
\be
A_\mu = -\frac{\delta {\cal L}}{\delta(\partial_\mu \beta(x))} = - \ol{L}\gamma_\mu\gamma_5 L +
 \left((\partial_\mu \phi)b - (\partial_\mu b)\phi\right)
\ee
Now add to the Lagrangian a piece ${\cal L}'(\phi,b)$ given by
\be
{\cal L}'(\phi,b) = c_\phi \phi + c_b b
\ee
Then $\partial^\mu A(x)\neq 0$; it becomes
\be
\partial^\mu A_\mu = - \frac{\delta {\cal L}}{\delta\beta} = (c_\phi b - c_b \phi)
\ee
There are, in general, certain non-zero matrix elements
of $A_\mu$, defined by
\ba
\sqrt{2p_0^b}\langle 0| A_\mu(x)|b\rangle&=& \i (p_\mu^b) {\mathrm e}^{-\i(p_\mu^b)x^\mu} \phi_c\nn\\
\sqrt{2p_0^\phi}\langle 0| A_\mu(x)|\phi\rangle&=& -\i (p_\mu^\phi) {\mathrm e}^{-\i(p_\mu^\phi)x^\mu} F_b
\ea
Differentiating, and using eq.~(A5) gives
\ba
c_\phi &=& m_b^2 \phi_c\nn\\
c_b &=& m_\phi^2 F_b
\ea
We rewrite ${\cal L}(\phi,b)$ in terms of fluctuating fields, that is $\phi\to (\phi_c + (\delta\phi))$
and $b\to (F_b + (\delta b))$, using the the notation $(\delta \phi)=\tilde{\phi}$ and
$(\delta b) = \tilde{b}$. We include all terms in $\phi_c$ and terms linear in $F_b$ - these
terms contain the essential points that we wish to make - because of the hypothesis that $F_b\ll \phi_c$
(note footnote F10). 
\ba
{\cal L}(\tilde{\phi},\tilde{b}) &=& \frac{\i}{2}\left(\ol{L}\gamma_\mu(\partial^\mu L) -
(\partial^\mu\ol{L})\gamma_\mu L\right) - m_L \ol{L}L\nn\\
&+&\frac{1}{2}\left((\partial_\mu\tilde{\phi})^2 - m_\phi^2\tilde{\phi}^2\right)
+\frac{1}{2}\left((\partial_\mu \tilde{b})^2 - m_b^2 \tilde{\phi}^2\right)\nn\\
&-& g\ol{L}L \tilde{\phi} - \i g \ol{L}\gamma_5 L \tilde{b} - \lambda\left(\tilde{\phi}^2 + \tilde{b}\right)^2\nn\\
&-&4\lambda \phi_c\tilde{\phi}\left(\tilde{\phi}^2 + \tilde{b}^2 + 2F_b\tilde{b}\right) -
4\lambda F_b\tilde{b}\left(\tilde{\phi}^2 + \tilde{b}^2\right)\nn\\
&+&(c_\phi - m_b^2\phi_c) \tilde{\phi} + (c_b - m_b^2 F_b)\tilde{b}
\ea
with
\ba
m_L&=& g\phi_c\nn\\
m_\phi^2 &=& (\mu^2 + 12\lambda \phi_c^2)\nn\\
m_b^2 &=& (\mu^2 + 4\lambda \phi_c^2)
\ea
${\cal L}(\tilde{\phi},\tilde{b})$ does not treat $\tilde{\phi}$ and $\tilde{b}$
symmetrically, because $F_b\neq \phi_c$. Setting the coefficients of the last
two terms in ${\cal L}(\tilde{\phi},\tilde{b})$ to zero
($\langle 0| {\cal L} |\tilde{b}\rangle = 0 = \langle 0|{\cal L}|\tilde{\phi}\rangle$),
gives
\ba
c_\phi &=& m_b^2 \phi_c,\;\;\;\; \mbox{as in eq. (A7)}\nn\\
c_b &=& m_b^2 F_b \;\;\;\; (=m_\phi^2 F_b \;\;\mbox{from eq. (A7)})
\ea
Consider the limiting situation in which ${\cal L}'$ vanishes \cite{ref23}. In the
usual Goldstone mode\cite{ref25}, $\phi_c\neq 0$. As $c_\phi$
approaches zero, $m_b^2$ approaches zero, and $\mu^2$ approaches $-4\lambda\phi_c^2<0$.
Then $m_\phi^2$ tends to $8\lambda\phi^2_c>0$. If $F_b\neq 0$ and $c_b$ approaches zero,
$m_\phi^2$ must approach zero. Then $\lambda$ approaches zero, and so $\mu^2$ approaches
zero through negative values. In other words, one appears to be just inside the
boundary for spontaneous symmetry breaking.\par
In the model of eq.~(1), the required value of $\lambda$ is non-zero
but is indeed a very small number, $10^{-13}$. A reason for $\lambda$
being such a small number has not been present before\cite{ref2,ref6}. The coefficient
of the $\phi^2 b^2$ interaction, $\lambda_b$, is much larger than $2\lambda$. There
is an explicit breaking of the symmetry. This results in a relatively large
mass for $m_b$, when the radiatively-corrected, tree-level potential for $\phi$
acquires a minimum at $\phi=\phi_c$.\par
A unitary transformation makes manifest\cite{ref18} the CP violation in the interactions
in ${\cal L}(\tilde{\phi},\tilde{b})$:
\be
L\to {\mathrm e}^{-\i \gamma_5 \frac{\alpha}{2}} L\;\;\;\;  \ol{L}\to \ol{L}{\mathrm e}^{-\i \gamma_5 \frac{\alpha}{2}}
\ee
with $\tan\alpha = (gF_b/m_L)$. This results in the following changed form for a part of 
${\cal L}(\tilde{\phi},\tilde{b})$,
\ba
&&\left(m_L \ol{L}L + \i g\ol{L}\gamma_5 L \tilde{b} + g\ol{L} L \tilde{\phi}\right)  \to\nn\\
&&\left(\tilde{m}_L \ol{L}L + g\left(\i\ol{L}\gamma_5 L (\cos\alpha) 
         + \ol{L} L (\sin\alpha)\right)\tilde{b}\right.\nn\\
&&+ g\left.\left(\ol{L}L(\cos\alpha) - \i \ol{L}\gamma_5 L (\sin\alpha)\right)\tilde{\phi}\right)
\ea
where $\tilde{m}_L = \sqrt{m_L^2 + (gF_b)^2}$.
CP violation occurs for $\tilde{\phi}$, as well as for $\tilde{b}$. (Note also the CP-violating
mixing term $(-8\lambda\phi_c F_b)\tilde{\phi}\tilde{b}$, in eq.~(A8).)
\clearpage
%
% figures
%
\section*{Figures}
\begin{figure}[h]
\begin{center}
\mbox{\epsfysize 15cm \epsffile{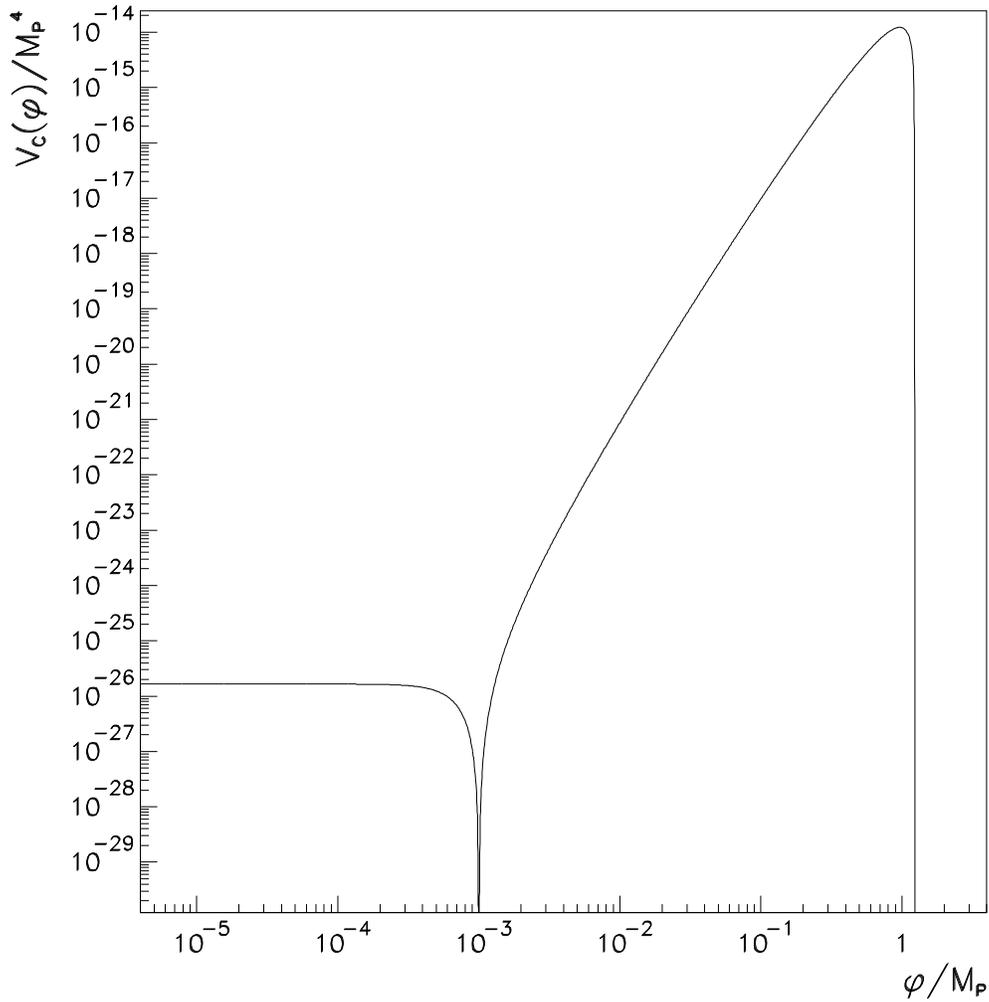}}
\caption{\small An example of the radiatively-corrected, tree-level potential $V_c(\phi)$ which
sets up inflation, as calculated by solving eqs.~(2,3) in the text. The maximimum is
at $\phi=\phi_m\protect\cong M_P\protect\cong 10^{19}$ GeV, and the minimum is at $\phi=\phi_c \protect\cong
10^{-3}M_P\protect\cong 10^{16}$ GeV. The inflaton mass is $m_\phi\protect\cong 5\times 10^{-10} M_P$.
(Addition of a constant renormalizes the calculated curve so that $V_c(\phi_c)=0$.)}
\end{center}
\end{figure}
\clearpage
\begin{figure}[t]
\begin{center}
\mbox{\epsfysize 15cm \epsffile{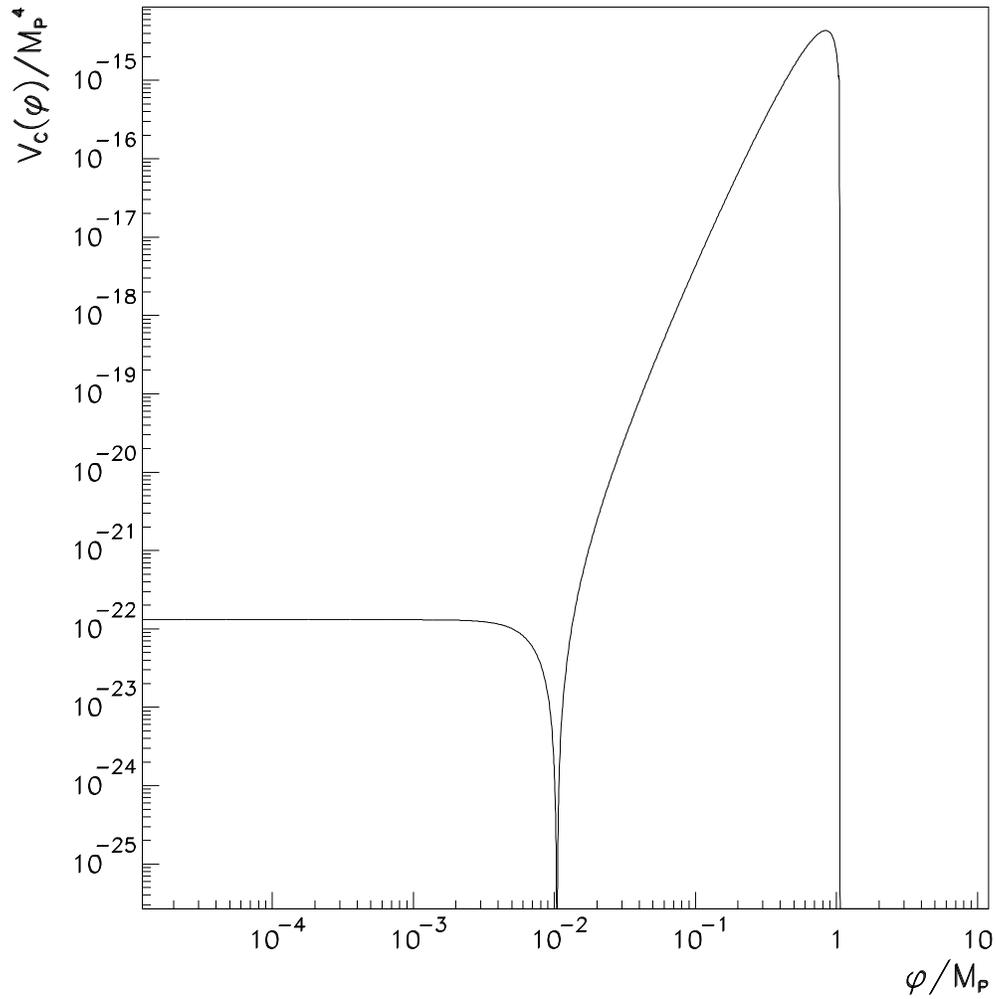}}
\end{center}
\caption{\small A second example of the radiatively-corrected, tree-level potential.
The maximum is at $\phi=\phi_m\protect\cong M_P$, and the minimum is at $\phi=\phi_c\protect\cong 10^{-2}M_P$.
The inflaton mass is $m_\phi\protect\cong 4\times 10^{-9}M_P$.}
\end{figure}
\clearpage
\begin{figure}[t]
\begin{center}
\mbox{\epsfysize 15cm \epsffile{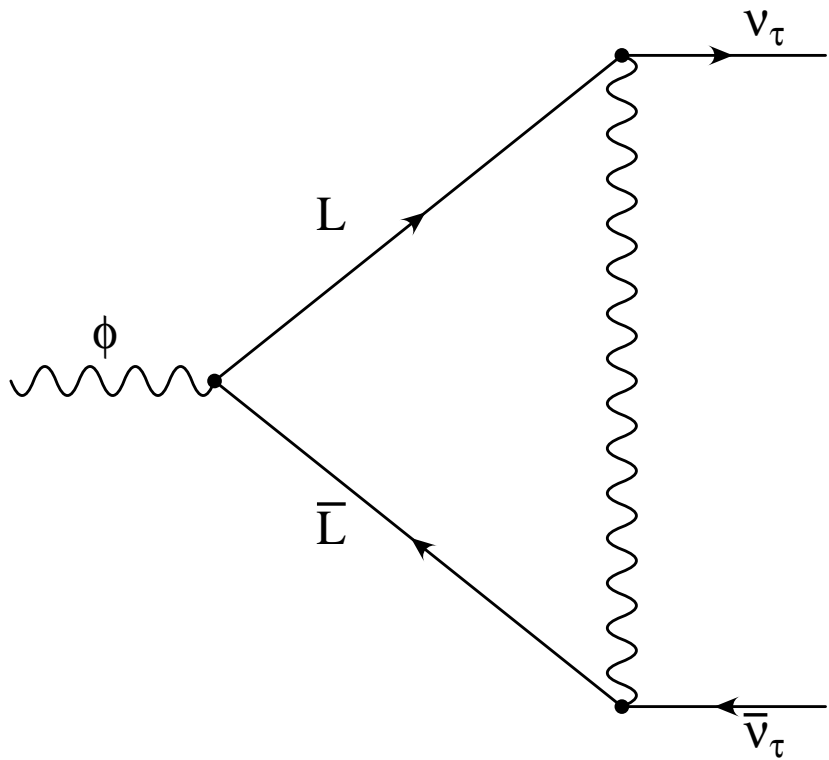}}
\caption {\small The process which gives rise to the eventual decay of massive quanta of the inflaton
field, via a virtual $L,\ol{L}$ pair, if $m_{\nu_\tau}\neq 0$.}
\end{center}
\end{figure}
\clearpage
\begin{figure}[t]
\begin{center}
\mbox{\epsfysize 15cm \epsffile{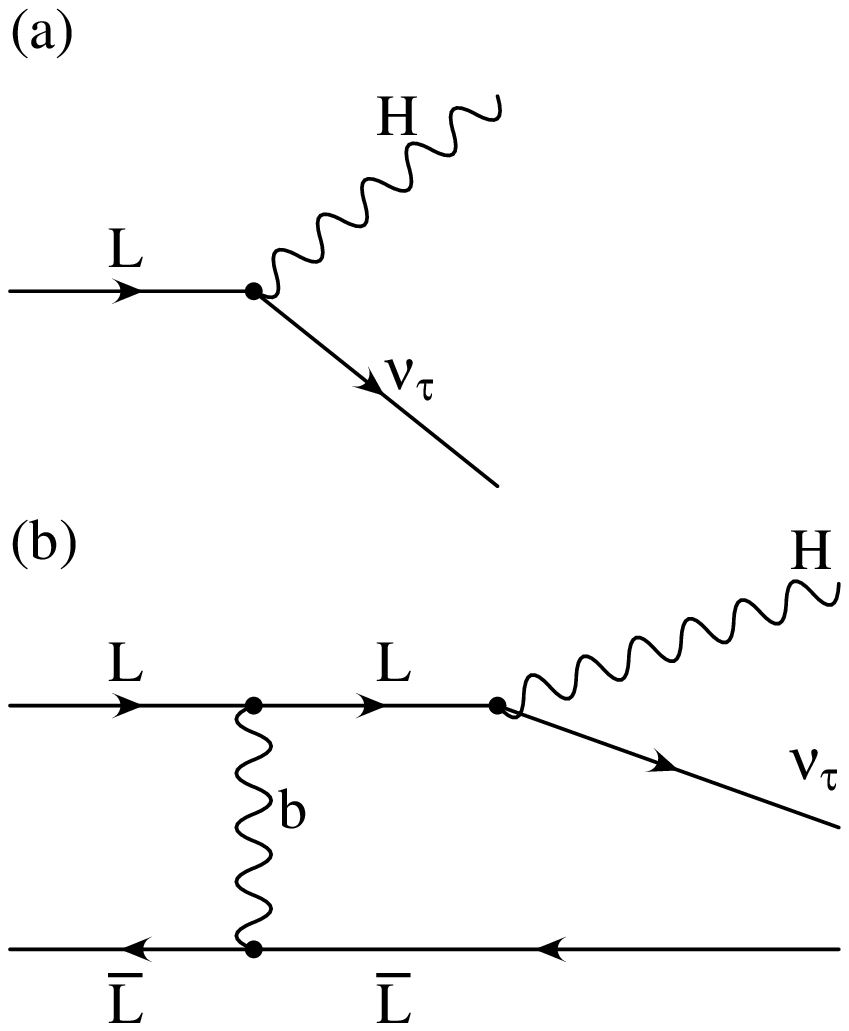}}
\caption{\small (a) The decay of an isolated $L$ (or $\protect\ol{L}$), via a mixing interaction to
a light neutrino with mass $m_{\protect\nu_\protect\tau}$.\protect\newline
(b) The decay of an $L$ considered to be in a quasi-bound, metastable state with an $\protect\ol{L}$,
which condition is brought about by interaction involving the exchange of $b$-quanta. The
process is not symmetric for $L$ decay and $\protect\ol{L}$ decay.}
\end{center}
\end{figure}
\end{document}